\documentclass[8.5pt]{article}
\usepackage[super,sort&compress,comma]{natbib} 
\usepackage{mhchem}
\usepackage{times,mathptmx}
\usepackage{sectsty}
\usepackage{balance}

\usepackage{graphicx} 
\usepackage{epstopdf}
\usepackage{lastpage}
\usepackage[format=plain,justification=raggedright,singlelinecheck=false,font=small,labelfont=bf,labelsep=space]{caption} 
\usepackage{fancyhdr}
\begin{document}

\thispagestyle{plain}
\fancypagestyle{plain}{
\renewcommand{\headrulewidth}{1pt}}
\renewcommand{\thefootnote}{\fnsymbol{footnote}}
\renewcommand\footnoterule{\vspace*{1pt}%
\hrule width 3.4in height 0.4pt \vspace*{5pt}} 
\setcounter{secnumdepth}{5}

\makeatletter 
\def\subsubsection{\@startsection{subsubsection}{3}{10pt}{-1.25ex plus -1ex minus -.1ex}{0ex plus 0ex}{\normalsize\bf}} 
\def\paragraph{\@startsection{paragraph}{4}{10pt}{-1.25ex plus -1ex minus -.1ex}{0ex plus 0ex}{\normalsize\textit}} 
\renewcommand\@biblabel[1]{#1} 
\renewcommand\@makefntext[1]%
{\noindent\makebox[0pt][r]{\@thefnmark\,}#1}
\makeatother 
\renewcommand{\figurename}{\small{Fig.}~}
\sectionfont{\large}
\subsectionfont{\normalsize} 

\fancyfoot{}
\fancyfoot[RO]{\footnotesize{\sffamily{1--\pageref{LastPage} ~\textbar \hspace{2pt}\thepage}}}
\fancyfoot[LE]{\footnotesize{\sffamily{\thepage~\textbar\hspace{3.45cm} 1--\pageref{LastPage}}}}
\fancyhead{}
\renewcommand{\headrulewidth}{1pt} 
\renewcommand{\footrulewidth}{1pt}
\setlength{\arrayrulewidth}{1pt}
\setlength{\columnsep}{6.5mm}
\setlength\bibsep{1pt}

\twocolumn[
 \begin{@twocolumnfalse}
\noindent\LARGE{\textbf{Ab initio Screening of a Sulfur Desorbed MoS$_2$ Photocatalyst for Nitrogen Fixation}}
\vspace{0.6cm}

\noindent\large{\textbf{Alhassan S. Yasin\textit{$^{a}$} and Nianqiang Wu\textit{$^{a}$} and Terence Musho,$^{\ast}$\textit{$^{a}$}}}\vspace{0.5cm}

\noindent\textit{\small{\textbf{Received Xth XXXXXXXXXX 20XX, Accepted Xth XXXXXXXXX 20XX\newline
First published on the web Xth XXXXXXXXXX 200X}}}

\noindent \textbf{\small{DOI: 10.1039/b000000x}}
\vspace{0.6cm}

\noindent \normalsize{The following study investigates the thermodynamic reaction barriers during nitrogen fixation for an inorganic sulfur desorbed photocatalyst Molybdenum disulfide surface. The design space is investigated using density functional theory (DFT) method within a space bounded by MMoS$_2$ M=Mo,Fe,Co. The discussion focuses on Heyrovsky type reactions along both the associative and dissociative pathway. A key insight into the roles of the inorganic and the balance between nitrogen and hydrogen affinity, providing evidence for an optimal material that minimizes the required over-potential. It is found that phases with a higher concentration of Mo face high reaction barriers involving nitrogen, where phases with higher concentrations of Fe and Co face high reaction barriers involving hydrogen species. In the absence of kinetic considerations, the best phase was predicted to be the 1T phase with a Mo$_{0.75}$Fe$_{0.25}$S$_{2}$ composition. This phase proved to have a balance of hydrogen and nitrogen affinity and follows the dissociative pathway, which can be evolved through non-thermal methods. }
\vspace{0.5cm}
 \end{@twocolumnfalse}
 ]

\footnotetext{\textit{$^{a}$~Department of Mechanical and Aerospace Engineering, West Virginia University, Morgantown, WV 26506-6106, USA. Fax: 304-293-6689; Tel: 304-293-3256; E-mail: tdmusho@mail.wvu.edu}}

\section{Introduction}
Essential part of biological development and adaptation on this planet is the conversion of nitrogen into ammonia (nitrogen fixation).~\cite{hoffman14,ferguson98,burris01,canfield10} This is due to the fact that nitrogen is needed for the utilization of biogenesis~\cite{burris01}, which is an enzyme-catalyzed process where substances are converted into more complex products for basic building blocks of living organisms.~\cite{hoffman14} Eventhough nitrogen is one of the most abundant elements on Earth, dominantly in the form of nitrogen gas (N$_{2}$) in the Earth’s atmosphere~\cite{jia14,mackay04}, living organisms can only utilize reduced forms of nitrogen. Organisms utilize the reduced form of nitrogen through several forms, for example: through (i) ammonia (NH$_{3}$) and/or nitrate (NO$_{3}^{-}$) fertilizer~\cite{canfield10,cheng08,thamdrup12}, (ii) utilization of released compounds during organic matter decomposition~\cite{hoffman14,burris01}, (iii) the conversion of atmospheric nitrogen by natural processes, such as lightening~\cite{gruber08}, and (iv) biological nitrogen fixation (BNF)~\cite{hoffman14}. 

This study aims to depict key aspects that characterize BNF processes for an alloyed sulfur desorbed photocatalytic ternary system MMoS$_2$ M=Mo,Fe,Co. The MoS$_2$ based sulfur desorbed system tends to decrease the coordination of the Mo sites, which in contrast to the desorption of sulfur atoms the surface would lack affinity for nitrogen species. Experimentally, the desorption of sulfur has been achieved during ultra-high vacuum (UHV) annealing~\cite{donarelli13} of this transition metal dichalcogenides(TMD). TMD or TMDC share similar properties to graphene, which makes it an advantageous photocatalyst~\cite{wang12}. From a morphology perspective TMDs form two-dimensional structures, similar to graphene~\cite{wang12,pacile08}. However, differences are realized when expresing TMDs and graphene's electronic properties. A commonly studied TMD and the focus of this study is a Molybdenite TMD, which is commonly refered to as MoS$_2$ monolayers. The optical band gap properties for MoS$_2$ range from 1.2eV to 1.8eV depending on the monolayer environment~\cite{kam82,kuc11,han11}. There are several experimental approaches to achieving the monolayer phase that include both mechanical and chemical exfoliation. It was found that during chemical exfoliation and a post anneal phase at UHV~\cite{donarelli13} that sulfur desorption is realized and sulfur vacancies are formed. With this experimental method being known, the approach taken is to study the photocatalytic properties of a desorbed MoS$_2$ monolayer. 

\subsection{Biological Nitrogen Fixation (BNF)}
In nature, the BNF process occurs naturally in soil by nitrogen-fixing bacteria that are affiliated with some plants. Biological nitrogen fixation is carried out by specific groups of prokaryotes, these organisms utilize the enzyme nitrogenases~\cite{raymond04,hoffman14} (enzymes that are produced by certain bacteria, and are responsible for the reduction of nitrogen to ammonia). The nitrogen molecule is comprised of two nitrogen atoms joined by a triple covalent bond, making the molecule highly inert~\cite{jia14,mackay04} and nonreactive. Microorganisms that fix nitrogen (nitrogenases) require 16 moles of adenosine triphosphate (ATP) to reduce each mole of nitrogen (approximately 5 eV)~\cite{hubbell98}. Reduction of atmospheric nitrogen takes place in nitrogenases by initially weakening the N-N bond by successive protonation until the dissociation barrier is low enough that the N-N bond breaks (later in the reaction sequence)~\cite{howalt13}. This particular reaction sequence is referred to as the associative mechanism. The energy required by the microorganisms to reduce nitrogen is obtained by oxidizing organic molecules. Meaning that non-photosynthetic living microorganisms must obtain these molecules from another organisms, while photosynthetic capable microorganisms use sugars produce by photosynthesis to obtain the essential energy to oxidize other organic molecules~\cite{hubbell98,buttel98}.

The active site in the enzyme is a cluster of FeMo$_{7}$S$_{9}$N, the FeMo-cofactor, with an electrochemical reaction:
$N_{2}+8(H^{+}+e^{-}) \rightarrow 2NH_{3}+H_{2}$.
These clusters have shown great stability in various configurations and sustainability in natural environments. Thus, most synthetically researched nitrogen fixation processes of transition metal configurations are based in some variation to molybdenum (Mo) structures.

A key insight into the nitrogen fixation process has been realized through a series of study on the biological nitrogen fixation processes. First, it was determined that nitrogenase of the BNF process is a two component system~\cite{bulen66,mortenson65,mortenson66} comprised of MoFe protein and the electron transfer Fe protein~\cite{winter76,hageman78,kim94}. Second, a reducing source and MgATP are needed for catalysis~\cite{bulen65,burns65,mortenson64,mortenson642} and that the Fe protein and MoFe protein associate and dissociate in a catalytic cycle involving single electron transfer and MgATP hydrolysis~\cite{hageman78,hoffman14}. Third, the MoFe protein contains two metal clusters, the iron-molybdenum cofactor (FeMo-co)~\cite{burgess90,shah77}, which provides an active site for substrate binding and reduction, and P-cluster (which involves the electron transfer from the Fe protein to FeMo-co)~\cite{peters95,kim92,ma96,lowe93}. Fourth, crystallographic structure are important for both Fe and MoFe proteins.~\cite{howard96,kim92,kim921,chan93,kim93} These catalytic advancements have greatly assisted in the reduction of reaction pressure and temperature of the Haber Process. However, no significant energy consumption reduction has been shown, thus yielding the catalytic mechanism to remain incomplete.

\subsection{Haber-Bosch Process}
A typical method of synthetically producing ammonia is through the Haber-Bosch process~\cite{gruber08,hoffman14}. This process essentially reduces nitrogen the same way as biological systems~\cite{vitousek97}, however, with much greater energy demand. In this process conversion of atmospheric nitrogen (N$_{2}$) to ammonia (NH$_{3}$) is done by reactions with hydrogen (H$_{2}$), coupled with metal catalysts (typically an iron based catalyst). The reaction takes place under high temperature and pressure~\cite{burgess96,eady96}, where some energy can be recuperated. During the process N and H gas molecules are heated to approximately 400 to 450 $^{\circ}$C, while continuously being kept at a pressure of 150 to 200 atm. Prior to the exposure to the catalyst, it is necessary to remove as much of the oxygen as possible to avoid oxidation of the catalyst. The dissociated N and H mixture is passed over a Fe-based catalyst to form ammonia. 
\begin{equation}
N_{2}(g)+ 3H_{2}(g) \rightleftharpoons 2NH_{3}(g)
\label{equ:1}
\end{equation}


In contrast to BNF, the Haber-Bosch process initially dissociates the N$_{2}$ bond on the first step and then protonates each nitrogen atom, referred to as the dissociative mechanism~\cite{howalt13}. The nitrogen and hydrogen atoms do not react until the strong N$_{2}$ triple bond and H$_{2}$ bond have been broken~\cite{skulason12}. Eventhough the reaction is reversible and an exothermic process, relatively high temperature and pressure are still needed to make the reaction evolve quickly~\cite{kozuch08}. However, this shifts the equilibrium point towards the reactants thus resulting in the lower conversion of ammonia~\cite{skulason12,howalt13}. To correct for this shift of the reactants equilibrium, high pressure is thus needed to shift the equilibrium in favor of the reactions products~\cite{skulason12}.


\section{Methodology}
There is a growing concerns and need for synthetically producing ammonia that will need to be addressed with the growing population. What this study aims to understand is the possibility of utilizing concepts of electrochemistry for ammonia production. Excellent insight into obstacles faced while developing catalytic materials for nitrogen reduction have been demonstrated in the past several years from a theoretical~\cite{howalt13,raymond04,hoffman14} and experimental~\cite{howalt13,heyrovsky27,tafel05} point of view, which provides a great starting point for further investigation on these catalytic materials. Previous studies also show that ammonia synthesis is very structure sensitive on metal surfaces and primarily occurs on surfaces steps of Fe and Ru~\cite{howalt13}, and potently could expect the associative mechanism to be even more structure sensitive. Thus, looking at the highly under coordinated MoS$_{2}$ structure, which provides a means to investigate this structure sensitive associative mechanism.

Development of density functional theory (DFT) allows for more feasible means of reliably modeling chemical reactions of expressed MoS$_{2}$ structure for investigation of the electrocatalytic production of ammonia. This study focuses on the highly under coordinated MoS$_{2}$ structures, which has shown oxidation reactions performed on transition metal nanoparticles consisting of metals at lower temperatures\cite{honkala05,howalt13}, and has shown to dramatically change the reactivity of inert metals~\cite{howalt13}. Also, due to the fact that various experimental work has demonstrated this structure to have good: light absorption, band gap variability, charge mobility, thermo stabilaty, and structure mobility~\cite{hubbell98,buttel98,jia14,mackay04,burgess90,shah77}. While previous studies primarily looked at step metal surfaces of MoS$_{2}$, this study performed calculations on consistent basis for configurations of MoS$_{2}$ and M$_{x}$Mo$_{y}$S$_{2}$ (M=Fe, Co) sub configurations. This allows the investigation of the reaction intermediates for the dissociative and associative mechanisms on MoS$_{2}$ structure for Mo, Co, and Fe (M$_{x}$Mo$_{y}$S$_{2}$ (M=Fe, Co)) transition metals, and develop stepped and closed packed metal surface relations. These calculations use statistical mechanics to construct Gibbs free energy diagrams for both reaction pathways based on entropy, zero point energy corrections, and vibrational calculations. These diagrams are then used to determine the lowest potential across the investigated metal configurations, which will make each part an exothermic reaction step.

\subsection{Nitrogen Reduction}
Nitrogen reduction is a product of oxidation and reduction reactions, the loss and gain of electrons in a chemical reaction. Biochemical cycling of C, H, and O are the foundations in which oxidation and reduction reactions take place~\cite{arp00} in chemical elements with which nitrogen reduction is most critically associated with. Were nitrogen valence range undergoes a biochemical cycling~\cite{arp00,ferguson98} from which nitrogen can lose all five of its outer shell electrons to surrounding elements to gain three electrons from other elements to complete all orbitals of its outer-most electron shells.

Potentially in this reduction, nitrogen takes two cycling forms, first N atom can lose outer shell electrons if the surrounding element has a higher electron affinity (to be more electronegative). For example, if the surrounding element is to be oxygen which is a more electronegative element (affinity for electrons is greater than nitrogen), thus N will eventually lose all five outer shell electrons and thus N can eventually become fully oxidized as nitrate (NO$_{3}^{-}$). In contrast, N can eventually fill all outer shell electron orbitals (gain three electrons) from elements such as H and C elements, which are less electronegative than N. With this gain of electrons, N can be fully reduced to ammonia (NH$_{3}$), and potentially N can be full to partially reduced in various organic compounds. Same as C, H, and O, the N cycling involves a zero valence form (N$_{2}$ gas), for which the charge of the seven protons in N's nucleus are balanced by seven electrons orbiting the nucleus by the two electrons of the inner electron shell and the five of the outer shell. However, the depicted N reduction is greatly idealized and can therefore contain many intermediate oxidation/reduction forms that N can assume that are not expressed but are discussed in great detail in references~\cite{arp00,ferguson98,galloway08}.

Furthermore, the oxidation/reduction reactions of N are mediated by biological, chemical, and physical factors~\cite{arp00}. These factors do not only influence the form that N takes, but also the flux and accumulation of nitrogen, as well as the nature and extent of the reactions in which N contributes. The importance of these mediating factors are expressed from a thermodynamic examination of N reduction, for which N exists in forms other than its most thermodynamically stable form (NO$_{3}^{-}$)~\cite{ferguson98, galloway08}. This indicates that external sources of energy that are directed by mediating chemical, physical and biological factors truly contribute to the possibility of driving nitrogen reduction.

\subsection{Photocatalytic}
The main issue of synthetically reducing nitrogen is the energy required for the process to successfully be implemented. Typically this energy is attained by burning fossil fuels that yield environmental pollution as discussed in previous sections. This prompted researchers to investigate better pathways for ammonia production through the mimicry of biological organisms, thus allowing various researchers to try and develop catalysts for this process. This yielded many researchers to investigate the mechanism of nitrogen fixation and produce photocatalytic and photoelectrochemical processes to utilize solar energy conversion. 

The idea is to allow photoreaction to be accelerated by the presence of a catalyst, meaning that the photocatalytic activity is dependent on the catalyst to create hole pairs (electron charge carriers), which generate free unpaired valence electrons that can contribute to secondary reactions. Typically, photocatalysts are transition metal oxides and/or semiconductors that have void energy region (band gap) where no energy levels are present to undergo an electron and hole recombination produced by photoactivation (light/photon absorption). When the energy of an absorbed photon is equal to or greater than that of the materials band gap, an electron becomes excited from the valence band (bottom of band gap) to the conduction band (top of band gap), thus producing a positive hole in the valence band. This excited electron and hole become recombined releasing heat from the energy gained by the photon exciting the surface of the material. The idea is to have a reaction that produces an oxidized product by reaction of generated holes with a reducing agent and to have a reduced product by the interaction of the excited electron and an oxidant. Simply-put, photoelectrochemical and photocatalytic processes rely on semiconductor materials to absorb sunlight and generate excited charge carriers that allow for reactions to take effect without any external electricity input. 

Thus yielding a photocatalytic and electrochemical investigation for sunlight-driven nitrogen fixation to reduce energy consumption, which will also provide flexibility in designing materials for nitrogen conversion. This work prompts to gain a crucial understanding of nitrogen to ammonia reaction process on the photocatalytic and electrochemical level and to develop means of approaching a sustainable solar driven conversion process of nitrogen to ammonia.

\subsection{Material Phase}
\begin{figure}[!h]
\includegraphics[angle=90,width=1\columnwidth]{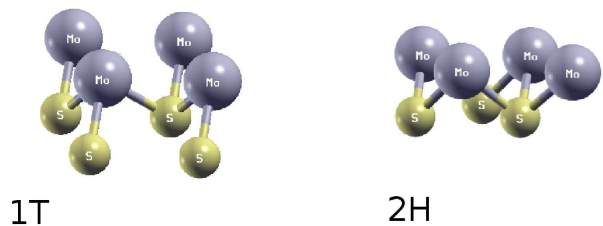} 
\caption{Illustration of the sulfur desorbed MMoS$_{2}$ 1T and 2H supercell structure. A 1T unit cell made up of 16 atoms 8 transition metal atoms and 8 S atoms, whereas the 2H unit cell is comprised of 18 total atoms 7 transition metal atoms and 11 S atoms. The transition metal atoms in the structures are alloyed with all combinations of Mo, Fe, Co atoms. }\hrule
\label{fig:UC}
\end{figure}

The material of interest is a MMoS$_{2}$ (Figure\ref{fig:UC}) that has received tremendous attention due to the earth-abundant composition and attractive optical, catalytic, and electronic properties, and high chemical stability. However, the focus of MoS$_2$ has mainly been focused on the hydrogen evolution reaction (HER)~\citep{wang16}. The bulk crystal is indirect gap semiconductor that has an energy gap of approximately 1.29 eV and is built up of van der Walls bonding of the S-Mo-S units. Due to their large surface areas and highly dense active sites along edges, MoS$_{2}$ are potentially promising material for electrochemical reaction studies. However, MoS$_{2}$ have proven to have poor conductivity, which has limited their electrochemical response. The fact that MoS$_{2}$ monolayers is known to have two phases, the trigonal prismatic (2H) and octahedral (1T), makes it a unique structure to investigate\citep{jiang15,li16}. The 2H phase has proven to be very stable, but with poor conductivity. However, the 1T phase is particularly favoring a stable structure at room temperature, but is metallic and better conductivity in nature.\citep{jiang15} This is why the combination of Fe and Co into the MoS$_{2}$ is being looked at in this study, to allow ideal properties to be demonstrated for distinct structures. If the properties of the two phases are combined in a monolayer structure, such as higher stability of the 2H phase and the high conductivity of the 1T phase~\citep{jiang15,lin14,li16}, this will result in the very favorable structure for N reduction. The resulting structure would have both large specific surface area and high charge transfer abilities that can be achieved simultaneously.

The structural transformation between semiconducting (2H) and metallic (1T) phases of MoS$_{2}$ has been a topic of interest in the past. These phase transitions (1T/2H) is made up of Mo and or S atomic glide planes that require a precursor intermediate phase ($\alpha$-phase)~\citep{lin14}. Also, the phase transition is found to be associated with movement between B and y-boundaries~\citep{lin14}. In a thermodynamic system, the phase transition is fundamental phenomena and of great technological importance in the material science research, due to the fact that properties of a material are able to be altered without the need on additional atoms into the system.

The MoS$_{2}$ crystal (Figure\ref{fig:UC}) structure is made of atomic layers stacked by van der Walls forces, each layer is comprised of strong in-plane binding of S-Mo-S triple atomic planes.\citep{lin14,li16} Depending on the arrangements of the S atom, the crystal structure can appear in two distinct symmetry(2H and 1T) as expressed by Figure\ref{fig:UC}, the phases can be easily converted to each other through an inner layer atomic plane glide, which involves a transverse displacement of one of the S-planes. The 2H phase demonstrates hexagonal lattice with a threefold symmetry with an atomic stacking sequence of ABA (S-M-S), whereas the 1T phase similarly shows atomic stacking sequence of ABA (S-M-S) however, the S-plane occupies the hollow center of the 2H hexagonal lattice.\citep{lin14,li16,jiang15} This S plane glide occupies the hollow center site of the 2H hexagon, thus resulting in a 2H$\rightarrow$1T phase transition (Figure\ref{fig:UC}). Due to weak interactions between these monolayers coupled with much stronger intralayer interactions, the formation of ultrathin crystals of MoS$_{2}$ by micromechanical cleavage technique has been demonstrated. Systematic studies of the evolution of the optical properties and electronic structure of ultrathin MoS$_{2}$ crystals have been demonstrated and shown to be as a function of layer numbers. To further modify the activity of the structure towards nitrogen, sulfur atoms have been desorbed from the top surface. This has been achieved experimentally using a UHV anneal process~\cite{donarelli13}.

\subsection{Material Design Space} 
Typically a good strategy to design new catalyst for nitrogen reduction is by combining components that lie from both the dissociative and associative side. Initial DFT calculations show that MoS$_{2}$ to favor the dissociative side and with the addition of Fe and or Co, which typically favor the associative mechanism, the material of interest takes the form of M$_{x}$Mo$_{y}$S$_{2}$ (M=Mo,Fe,Co). The main structural modification that was done is the substitution of the non-organic ion with various combinations in all possible positions in the unit cell (MMoS$_{2}$, {M = Mo,Co,Fe}). The 8 (1T) and 7 (2H) atomic positions that were modified are clearly shown in Figure~\ref{fig:UC} of the unit cell for the 1T and 2H structures. Every configuration combination of the three elements (Mo, Co, Fe) is implemented and tested in the metal positions expressed in Figure~\ref{fig:UC} of the unit cell. These configurations are classified under one element case (pure Mo, Co, Fe structure), two element cases (combination of Co and Fe, Co and Mo, Fe and Mo), and three element cases (combination of Co and Fe and Mo). 

It is noted that while the MoS$_{2}$ material is coupled with Fe and Co, all combination of the three elements are and will be evaluated in this study to better vary the mechanism of interest for each distinct adsorption species (N$_{2}$H$_{x}$, NH$_{x}$). For cases of elemental combination (two element and three element cases), there are different configurations possible for same elemental ratios. This makes a total of 6561 (3 elements for 8 positions 3$^{8}$) unique configurations for 1T and 2187 (3 elements for 7 positions 3$^{7}$) unique configurations for 2H structures. Only the best configurations (lowest energy of the relaxed structure) for each distinct structure will be evaluated for the electrochemical reactions. For example, configuration of 1T structure of 1 atom Co and 7 atom Mo (Co1Mo7)can have 8 distinct configurations of Co being in any of the 8 metal positions, so only the lowest energy structure after the relaxation will be implemented in the electrochemical reactions, and this is the same for every other configuration for 1T and 2H (Co2Mo6, Co3Mo5, ...). 

Also, its is noted that for each configuration there are various operations that had to be done in sequence to obtain proper data, this resulted in a total of 45 configurations for 1T and 36 configurations for 2H that had the lowest energy structure for a specific configuration. The reader should note that this is a tremendous amount of computational and analytically challenging process. This is an area that needs to be investigated from a statistical point of view. Then only the 45 (1T) and 36 (2H) unique structures went on to be implemented in the electrochemical reactions. This will significantly allow every configuration of 1T and 2H structure to be explored and give the best configuration of MMoS$_{2}$, {M = Mo,Co,Fe} structure as a means of nitrogen reduction catalyst.

\subsection{Associative and Dissociative Reaction Pathways}
In the process of electrochemically forming ammonia, it is conventional to reference the source of protons and electrons to model the anode reaction
\begin{equation}
H_{2} \rightleftharpoons 2(H^{+}+e^{-})
\label{equ:2}
\end{equation}
First protons are introduced into the proton conducting electrolyte to sustain the equilibrium and diffuse into the cathode, while an external circuit is used to transport electrons to the cathode side through a wire. At this point, a nitrogen molecule will react with surrounding protons and electrons at the cathode as expressed by the following reaction
\begin{equation}
N_{2}+6(H^{+}+e^{-}) \rightarrow 2NH_{3},
\label{equ:3}
\end{equation}
to form ammonia at the catalytically active site, thus the overall electrochemical reaction is denoted by reaction (\ref{equ:1}).

Theoretically, the reaction can take the form of two different possible types of pathways in which there are two different possible types of mechanism for each pathway in order to synthesize ammonia electrochemically. The respected pathways are the associative and the dissociative pathway, where the possibility of either adsorbed N$_{2}$H$_{x}$ or NH$_{x}$ species can be hydrogenated. These pathways correspond to the Tafel-type mechanism and the Heyrovsky-type mechanism.~\cite{heyrovsky27} Solvated protons from the solution first adsorb~\cite{tafel05} on the surface and combine with electrons, then the hydrogen adatoms react with the adsorbed species(N$_{2}$H$_{x}$ or NH$_{x}$) in the Tafel-type mechanism. This type of mechanism can only have an indirect effect through interchangeable concentrations of the reactants.~\cite{heyrovsky27} So due to the fact that this study focuses on room temperature processes and also the activation barriers for Tafel-type reactions are about 1 eV or higher for most transition metal surfaces,~\cite{honkala05,wang11} this type of mechanism will most likely prove to be very slow. Also, this type of mechanism requires hydrogenation steps~\cite{wang11} of the reaction barriers to be overcome, and thus will also require higher temperature as well to drive the process forward. This is due to the requirement of the reaction to merge proton and electrons to form hydrogen adatom on the surface first.~\cite{tafel05} Thus the process will therefore either go through an associative and or dissociative Heyrovsky-type reaction. 

In Heyrovsky-type of reaction~\cite{heyrovsky27}, the adsorbed species are directly protonated so that a coordinate bond to the proton and the species are formed. Molecules from the electrolyte get directly attached by protons and then electrons from the surface merge with the protons to form a hydrogen bonded to the molecule. Also, by applying a bias in the latter cases of the mechanism, a thermochemical barrier can be directly affected. Initially, this study considers the possibility of the reaction to take an associative Heyrovsky mechanism, similar to that of the mechanism for the BNF, where the N-N bond is initially weakened by successive protonations until the dissociation barrier is low enough so that the N-N bond can be broken later in the reaction. For the Heyrovsky mechanism, nitrogen molecule is first attached to the surface and is then protonated before N-N bond dissociates. In the bellow equations of the associative Heyrovsky mechanism (asterisk '*', denotes a site on the surface):
\begin{equation}
N_{2}(g)+6(H^{+}+e^{-})+* \rightleftharpoons N_{2}*+6(H^{+}+e^{-})
\label{equ:4}
\end{equation}
\begin{equation}
N_{2}*+6(H^{+}+e^{-}) \rightleftharpoons N_{2}H*+5(H^{+}+e^{-})
\label{equ:5}
\end{equation}
\begin{equation}
N_{2}H*+5(H^{+}+e^{-}) \rightleftharpoons N_{2}H_{2}*+4(H^{+}+e^{-})
\label{equ:6}
\end{equation}
\begin{equation}
N_{2}H_{2}*+4(H^{+}+e^{-}) \rightleftharpoons N_{2}H_{3}*+3(H^{+}+e^{-})
\label{equ:7}
\end{equation}
\begin{equation}
N_{2}H_{3}*+3(H^{+}+e^{-})+* \rightleftharpoons 2NH_{2}*+2(H^{+}+e^{-})
\label{equ:8}
\end{equation}
\begin{equation}
2NH_{2}*+2(H^{+}+e^{-}) \rightleftharpoons NH_{3}*+NH_{2}*+(H^{+}+e^{-})
\label{equ:9}
\end{equation}
\begin{equation}
NH_{3}*+NH_{2}*+(H^{+}+e^{-}) \rightleftharpoons 2NH_{3}*
\label{equ:10}
\end{equation}
\begin{equation}
2NH_{3}*\rightleftharpoons NH_{3}*+NH_{3}(g)+*
\label{equ:11}
\end{equation}
\begin{equation}
NH_{3}*+NH_{3}(g)\rightleftharpoons 2NH_{3}(g)+*
\label{equ:12}
\end{equation}
The addition of the fourth H to the N$_{2}$H$_{3}$* molecule weakens the N-N bond to readily dissociates molecule into the species (NH$_{x}$) on the surface. Also, there is a possibility of reaction (\ref{equ:7}) to split into NH and NH$_{2}$ on the surface and has been observed on some metals.\cite{howalt13}

The second mechanism, dissociated Heyrovsky mechanism, is also considered and compared to the associative mechanism. Where in this mechanism the nitrogen molecule is initially dissociated on the surface, followed by subsequent protonation by direct attachment of protons:
\begin{equation}
N_{2}(g)+6(H^{+}+e^{-})+* \rightleftharpoons N_{2}*+6(H^{+}+e^{-})
\label{equ:13}
\end{equation}
\begin{equation}
N_{2}*+6(H^{+}+e^{-})+* \rightleftharpoons 2N*+6(H^{+}+e^{-})
\label{equ:14}
\end{equation}
\begin{equation}
2N*+6(H^{+}+e^{-}) \rightleftharpoons NH*+N*+5(H^{+}+e^{-})
\label{equ:15}
\end{equation}
\begin{equation}
NH*+N*+5(H^{+}+e^{-}) \rightleftharpoons NH_{2}*+N*+4(H^{+}+e^{-})
\label{equ:16}
\end{equation}
\begin{equation}
NH_{2}*+N*+4(H^{+}+e^{-}) \rightleftharpoons NH_{3}*+N*+3(H^{+}+e^{-})
\label{equ:17}
\end{equation}
\begin{equation}
NH_{3}*+N*+3(H^{+}+e^{-}) \rightleftharpoons NH_{3}*+NH*+2(H^{+}+e^{-})
\label{equ:18}
\end{equation}
\begin{equation}
NH_{3}*+NH*+2(H^{+}+e^{-}) \rightleftharpoons NH_{3}*+NH_{2}*+(H^{+}+e^{-})
\label{equ:19}
\end{equation}
\begin{equation}
NH_{3}*+NH_{2}*+(H^{+}+e^{-}) \rightleftharpoons 2NH_{3}*
\label{equ:20}
\end{equation}
\begin{equation}
2NH_{3}* \rightleftharpoons NH_{3}*+NH_{3}(g)+*
\label{equ:21}
\end{equation}
\begin{equation}
NH_{3}(g)+NH_{3}* \rightleftharpoons 2NH_{3}(g)+*
\label{equ:21}
\end{equation}

\subsection{Electrochemical Reaction}
For a deep understanding of the reactions, free energy correction is needed to be determined and included in the analysis for each reaction intermediate. From DFT calculations, a reasonable approximation to the free energy for the adsorbed species relative to the gas phase molecular N and H can be obtained from the expression:
\begin{equation}
\Delta G = \Delta E + \Delta E_{ZPE} - T\Delta S,
\label{equ:22}
\end{equation}
where $\Delta$E$_{ZPE}$ is the reaction zero point energy and $\Delta$S is the reaction entropy. In this study, only the ZPE is considered for the gas phases.


The applied reference potential driving the electrochemical reaction is set to be that of the standard hydrogen electrode (SHE), which this study also takes into account in addition to zero point energy and entropy. By using the computational SHE, this study is able to include the effect of the potential on the expressed reactions for surface sites. The standard hydrogen electrode has given great incite to describe numerous electrochemical reactions, such as trends in CO$_{2}$ (carbon dioxide), nitrogen, and oxygen reduction. 

Thus this study uses the standard hydrogen electrode as the reference potential, which expresses the free energy per H (chemical potential of (H$^{+}$+e$^{-}$) as related to that of $\frac{1}{2}$H$_{2}$(g), which is equation (\ref{equ:2}) in equilibrium. This implies that the pH = 0, the potential is that of U = 0 V relative to the SHE, and a pressure of 1 bar of H$_{2}$ in gas phase at 298 K, thus reaction (\ref{equ:1})'s free energy is equal to that of net reactions of (\ref{equ:4})-(\ref{equ:12}) or (\ref{equ:13})-(\ref{equ:21}) at an electrode. Also, it is noted that all presented calculations in this study are for that of pH equal to 0.

\subsection{Computational Details}
Ground state thermodynamic properties were predicted for thermodynamic steps that helped this study analyze various configuration reaction steps by means of density functional theory (DFT) approach.~\cite{qe} DFT calculations used functional form of the pseudo-wave function that was based on Perdew-Burke-Ernzerhof (PBE) exchange-correlation function at potentials with a cut-off wave function energy of 1224 eV (90 Ry), which provided most accurate and stable for the intended unit cell. Also, a pseudized wave function allowed the reduction of computational expense for which this study greatly benefited due to the various configurational combinations this structure had that this study hoped to explore. In addition, computational expense was greatly reduced when only a single primitive cell was simulated for each configuration. Monkhorst-Pack with a k-point mesh sampling 2x2x2 grid with an offset of 1/2,1/2,1/2 and a 7 {\r{A}} vacuum around the mono-layer has been applied. Van der Waals correction term~\cite{grimme06,barone09} was incorporated to account for the Van der Waals interaction, which allowed some empiricism into the calculation. Scaling parameters were specified to be 0.7 and cut-off radius for the dispersion interaction was 900 {\r{A}}. 

When solving electronic density self-consistently, adsorbates (N, H, NH, etc.) sitting on the structure and unit cell geometries were relaxed to a relative total energy less than 1x10$^{-10}$ and overall cell pressure of less than 0.5kBar. The unit cell was relaxed in the vacuum before any absorbates were implemented in the structure, then once relaxation of the structure was done the atoms were kept fixed so that the adsorbates were allowed to relax on the surface of the structure in the 7 {\r{A}} vacuum. The reader should note that pure DFT predictions of energetics, band gap, etc. are often under predicted due to the over-analyticity of the functionals and exchange-correlation terms. Therefore, the reported energies in this study should not be used as absolutes but used to study the trends for various structures.

\section{Results and Discussion}
Adsorption energies of all reaction states in (\ref{equ:4})-(\ref{equ:21}) are calculated for various combinations of M$_{x}$Mo$_{y}$S$_{2}$ (M=Fe, Co) in a 1T and 2H structure. Results are used to estimate the free energy change in elementary reactions (\ref{equ:4})-(\ref{equ:12}) for the associative mechanism and (\ref{equ:13})-(\ref{equ:21}) for the dissociative mechanism. The theoretical over-potential needed for reactions to take effect is approximated using free energy difference of elementary steps. Linear approximating relations for adsorption energies that correspond to adsorption species (N$_{2}$H$_{x}$, NH$_{x}$) assist in depicting trends in the catalytic activity of the implemented transition metals (Mo, Co, Fe). Its known that adsorption energies of simple hydrogen containing species such as N$_{2}$H$_{x}$, NH$_{x}$ depend linearly on the adsorption energy of the nitrogen atom. This paper establishes linear relations for N$_{2}$H$_{x}$, NH$_{x}$ species using absorption energies of N adatom and also express the relation of metal transition linear scheme for the expressed structure (1T, 2H). 

\subsection{Adsorption Sites}
\begin{figure}[!h]
\includegraphics[angle=90,width=1\columnwidth]{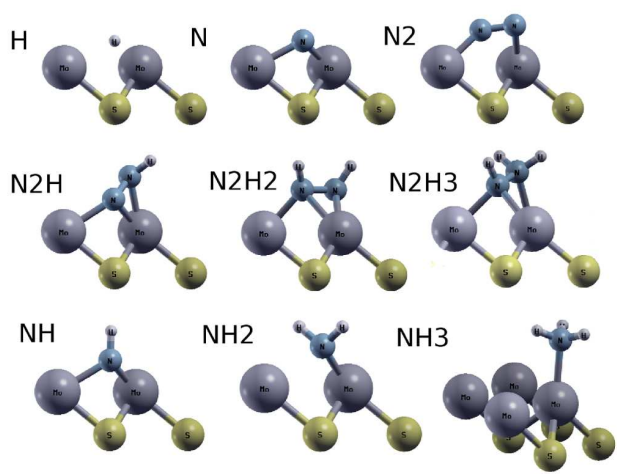} 
\caption{Absorption sites of the MoS$_{2}$ 1T structure. The absorption sites presented for the MoS$_{2}$ 1T structure is only used for the adsorption types and not for actual bonding for all other metals in this study. This means that the species of NH$_{x}$ could bind more strongly to Mo, Fe, Co structure configurations studied. }\hrule
\label{fig:ads}
\end{figure}

Adsorption sites illustrated in Figure\ref{fig:ads} of MoS$_{2}$ are used to demonstrate characteristics of adsorption states for all metals studied in this report because these sites have very similar trends; however, this is not to say that the binding energy is the same for different structures. It's sufficient to say that combinations of M$_{x}$Mo$_{y}$S$_{2}$ (M=Fe, Co) 1T and 2H structures produce very small variations in adsorption sites. These sites are usually classified as adsorption sites of hollow, bridge, and on top, and typically each classification will be slightly geometrically different. Thus each structure (M$_{x}$Mo$_{y}$S$_{2}$ (M=Fe, Co)) will yield unique electronic properties that can be analyzed and structures that prove more stable can be found and investigated.

Figure\ref{fig:ads} demonstrates these adsorption sites on MoS$_{2}$ monolayer, first hydrogen molecule proved to be quite unstable on the surface and tended to lie on the crystal surface. Thus hydrogen proved to have a bonding site of bridge classification; however, a very small difference of energies was found between adsorption sites. Whereas nitrogen adsorbed to the surface on a hollow site and preferred to bind in a di-sigma (strong covalent bond, formed by overlapping between atomic orbitals) bond on the surface with two metal atoms, thus proving edge sites to be the most stable. Similar in the absorption of N$_{2}$, however it is noted that N$_{2}$ molecule proved to have few stable adsorption configurations, due to the strong N-N bond.

The NH molecule binds to hollow sites, NH$_{2}$ to bridge sites, and NH$_{3}$ on-top sites and all proved to be stable structures (Figure~\ref{fig:ads} bottom row). However, in the case of N$_{2}$H, N$_{2}$H$_{2}$, and N$_{2}$H$_{3}$ species (Figure\ref{fig:ads} middle row), they typicality preferred to bond in a bridge site, and each nitrogen atom bonded to a metal atom similarly to the di-sigma bonding expressed for N and N$_{2}$ species (Figure\ref{fig:ads} top row). Based on the absorption sites, species' configuration orientation changed slightly as more hydrogen atoms are implemented in the surface species. Weakening of the N-N bond is demonstrated when looking carefully at Figure\ref{fig:ads}, as a visual representation of the internal bonding length dramatically increases as more hydrogen atoms are implemented in the species. It is very apparent when looking at NH$_{3}$ absorbed on the structure, were the internal bonding length is much greater to that of other species. Also, it is very apparent that nitrogen atom becomes further away from the metal atom as more hydrogen atoms become bonded to the respected nitrogen atom, this is demonstrated for BNF for the associative mechanism and shows the weakening of bonds between nitrogen and metal atoms.

\subsection{Ammonia Formation on Surface}
\begin{figure*}[!ht]
\begin{center}
\textbf{\begin{Large}2H\end{Large}}\hspace{85mm}\textbf{\begin{Large}1T\end{Large}} 
\end{center}
\includegraphics[width=0.5\columnwidth,angle=90]{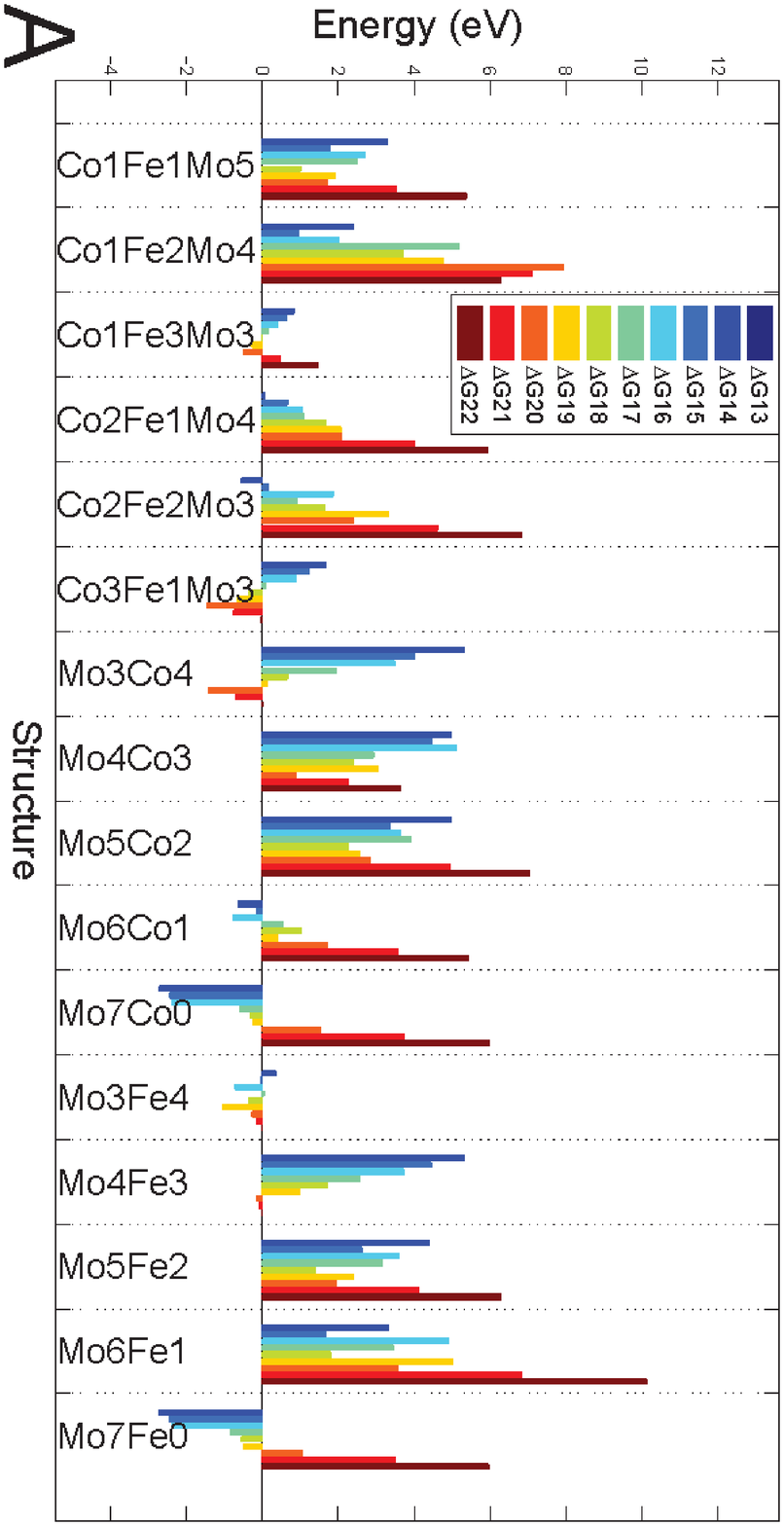}
\includegraphics[width=0.5\columnwidth,angle=90]{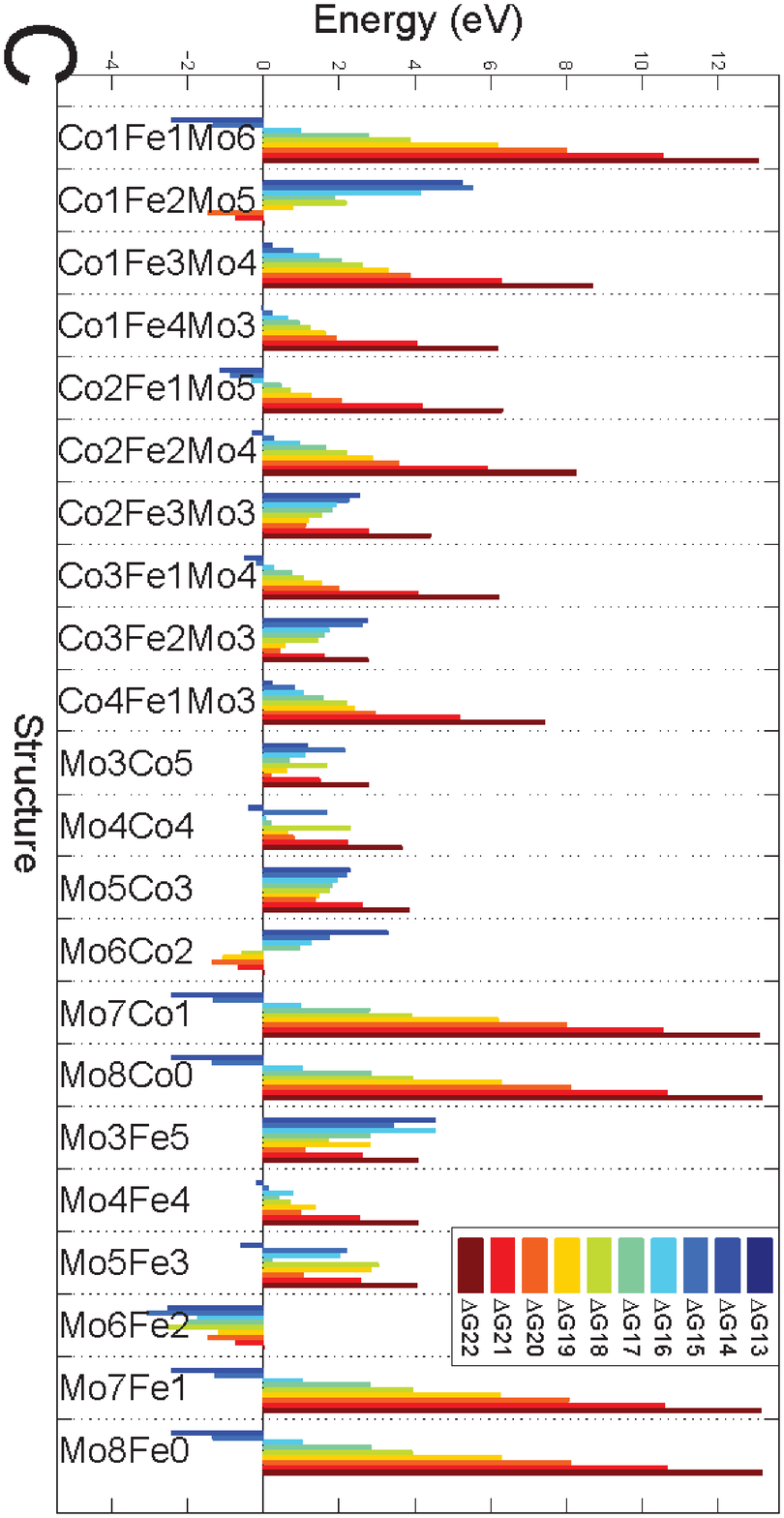} \\
\includegraphics[width=0.5\columnwidth,angle=90]{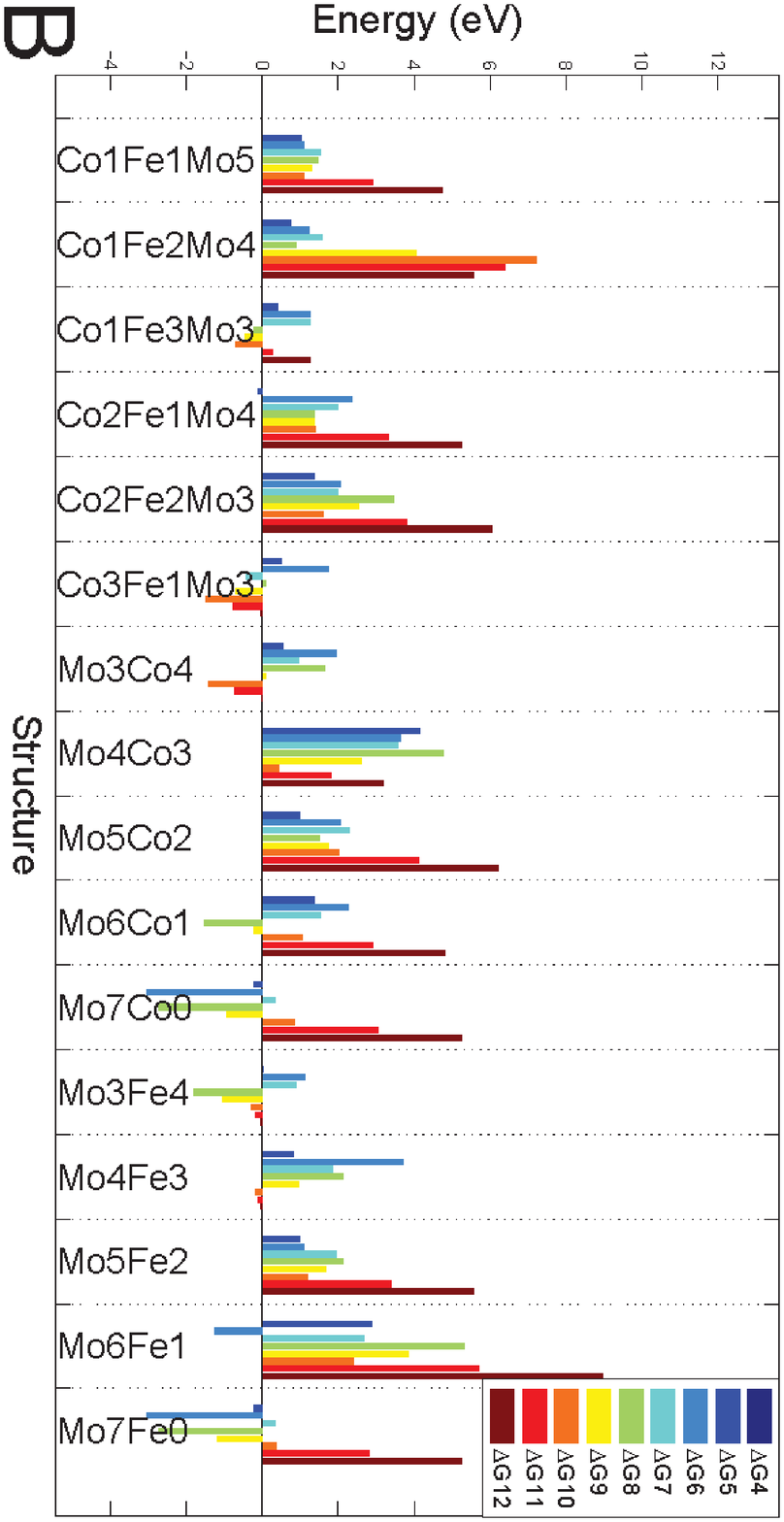}
\includegraphics[width=0.5\columnwidth,angle=90]{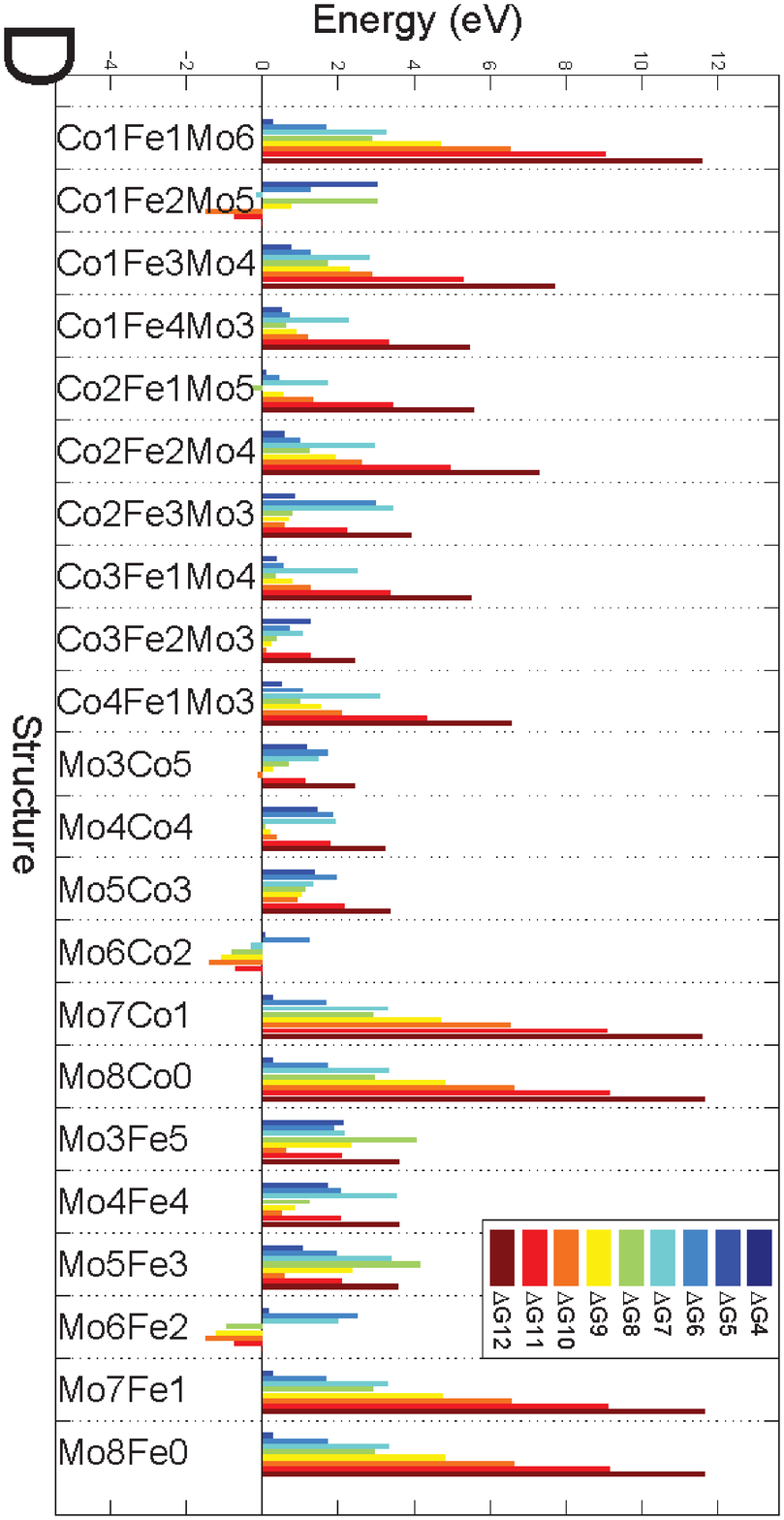}
\caption{Cumulative free energy for the dissociative (A,C) and associative (B,D) mechanism on both the 2H and 1T morphology. Energies are attained from DFT calculations for an electrolyte with pH=0 at 300 K. The energies correspond to reaction steps (\ref{equ:4})-(\ref{equ:12}) for the associative mechanism and (\ref{equ:13})-(\ref{equ:21}) for the dissociative mechanism relative. Each grouping can be associated with the potential-energy curve for successive reactions. The best performing structure and morphology is (Mo$_6$Fe$_2$)S$_2$ 1T phase (C).}\hrule
\label{fig:1T_step}
\end{figure*}

For each surface composition, two morphologies (1T,2H) with several adsorbate species that conform to the two reaction pathways were investigated to predict the electrochemical reaction of each step for different mechanism configurations. Figure\ref{fig:1T_step} illustrate the electrochemical reaction which refers to the free energy of reaction steps (\ref{equ:4})-(\ref{equ:12}) for the associative mechanism and (\ref{equ:13})-(\ref{equ:21}) for the dissociative mechanism of 1T and 2H structures.

DFT calculations show that a N$_{2}$ molecule binds to all surfaces of 1T and 2H structures with an adsorption energy that is always slightly more negative when compared to all other reaction steps, (Figure\ref{fig:1T_step}(A,C) dissociative ${\Delta}$G14 and Figure\ref{fig:1T_step}(B,D) associative  ${\Delta}$G5). The large loss in enthalpy in going from gas phase N$_{2}$ to a surface bonded molecule, this corresponds to a slightly negative free energy under ambient conditions for the 1T and 2H structures. For the most part, the associative mechanism for both structures (1T and 2H) tend to have the potential determining steps (largest positive step to overcome) in the reduction to form ammonia, this is the addition of one hydrogen atom to transition from N$_{2}$ to N$_{2}$H (Figure\ref{fig:1T_step} from ${\Delta}$G5 to ${\Delta}$G6), and addition of one hydrogen atom to go from NH$_{2}$ NH$_{3}$ (Figure\ref{fig:1T_step} from ${\Delta}$G9 to ${\Delta}$G10). This hydrogenation step corresponds to a positive step that is needed to overcome from free energy formation of N$_{2}$ and NH$_{2}$ on the surface, which is also deemed as a large energy barrier that is associated with the binding of hydrogen from the gas phase.

Typically the addition of hydrogen atoms promotes a positive step in free energy for most surfaces investigated and the first hydrogen step (from ${\Delta}$G5 to ${\Delta}$G6) appears to be the most uphill in free energy for all hydrogen steps. However, the over-potential in the dissociative mechanism for both morphologies (1T and 2H) is most significant when initially trying to break the N-N bonding going from N$_{2}$ to N (Figure\ref{fig:1T_step} from ${\Delta}$G14 to ${\Delta}$G15), this is very likely to occur due to N-N bond which binds in a di-sigma bond that is a very strong valent bond formed by overlapping between atomic orbitals that are very hard to overcome. Even thou N$_{2}$ binds to the surface with a negative adsorption energy from the significant loss in entropy from the gas phase, this is not enough to overcome this initial potential to brake N-N bonding without applying an external potential. It is also noted that for both the associative and dissociative mechanism of 1T and 2H structures (Figure\ref{fig:1T_step}) the last three steps (associative(B,D): ${\Delta}$G10 to ${\Delta}$G11 to ${\Delta}$G12 and dissociative(A,C): ${\Delta}$G20 to ${\Delta}$G21 to ${\Delta}$G22) are typically positive steps in free energy associated with the reduction of NH$_{3}$ on the surface to NH$_{3(g)}$ this is due to a large gain of entropy going from surface bounded molecule to a gas phase.



\subsection{Adsorption of N$_{2}$H$_{x}$ and NH$_{x}$ Species on Fe and Co Configurations}
\begin{figure*}[!ht]
\includegraphics[width=2.0\columnwidth]{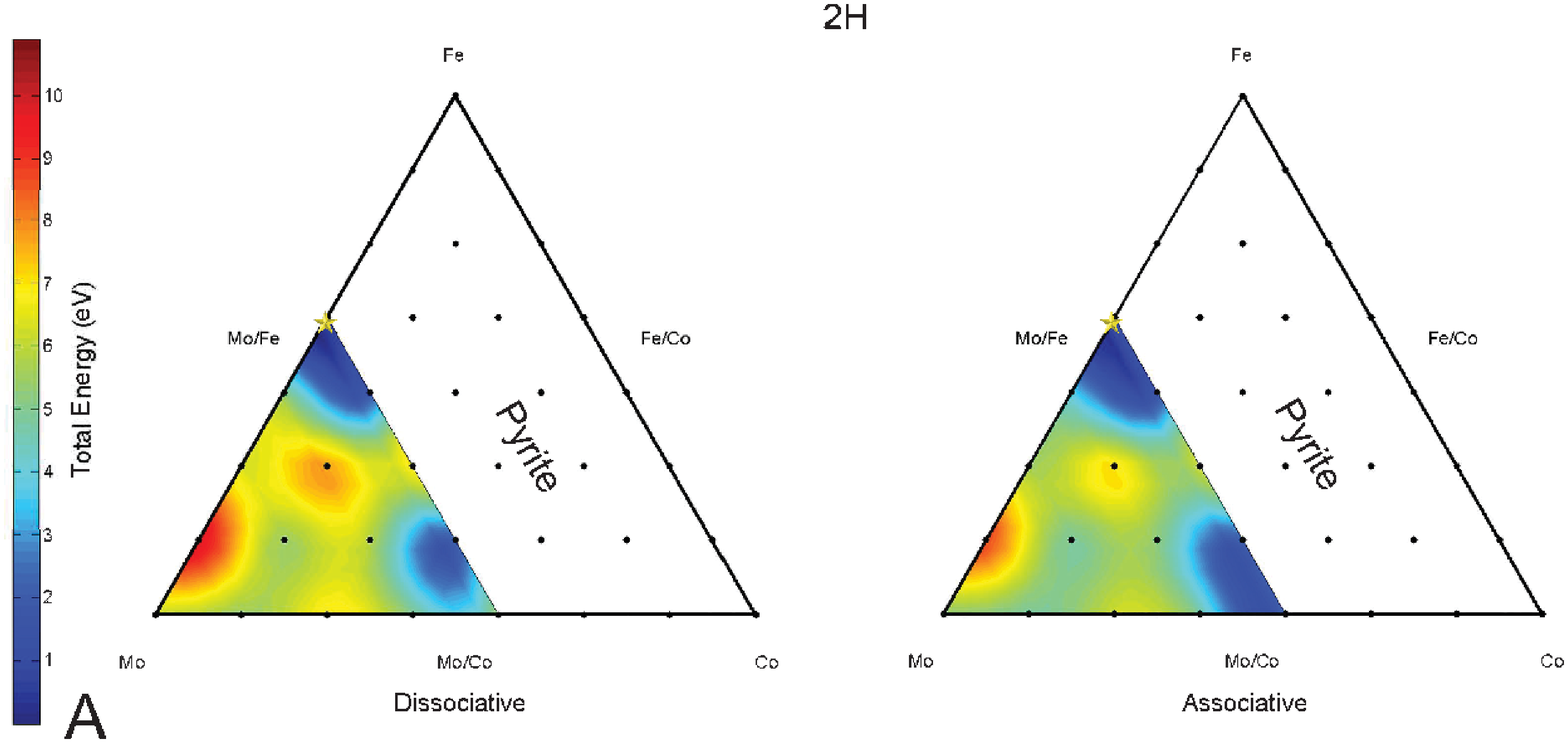}\\
\includegraphics[width=2.0\columnwidth]{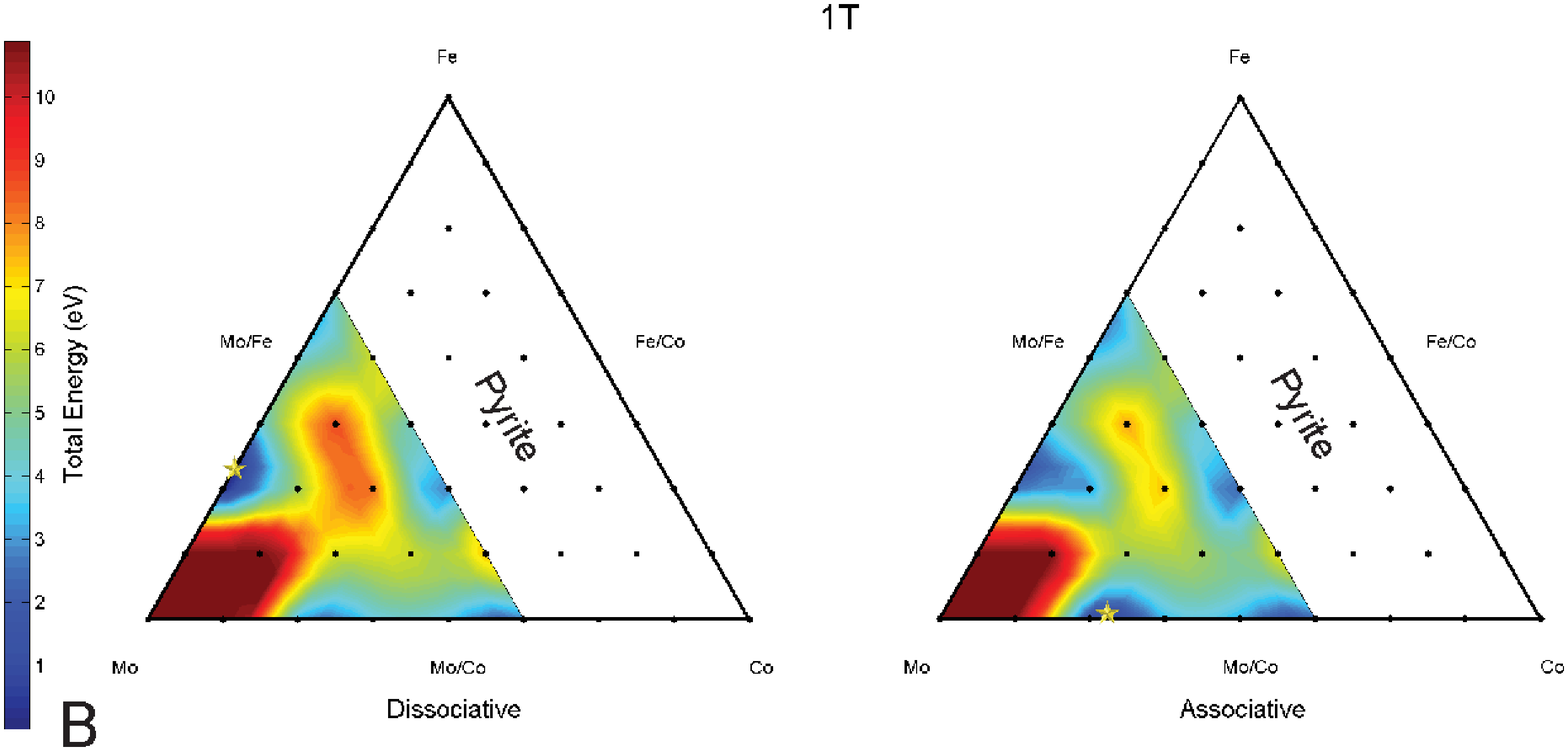} 
\caption{Ternary plots of the maximum reaction barrier for all successive reactions for the three inorganic components and the 2H (top) and 1T (bottom) morphologies. The contour colors corresponds to the energy of the most difficult step to overcome for reaction steps (\ref{equ:4})-(\ref{equ:12}) for the associative mechanism and (\ref{equ:13})-(\ref{equ:21}) for the dissociative mechanism. The white region of ternary plots corresponds to pyrite structure formation, which was not considered. The star in the figures represents the best possible configuration for the given pathway and morphology. The best points are tabulated in Table~\ref{table:best_points}. }\hrule
\label{fig:1T_turn_max}
\end{figure*}

The associative mechanism initially relies on absorbed configuration steps of N$_{2}$H$_{x}$ molecules whereas the dissociative mechanism initially relies on absorbed configuration steps of NH$_{x}$ molecules. The associated Gibbs energies of these reactions depends on both the morphology and composition of the surface. In this study stable configurations of the surface, which is determined based on the formation energy predictions, are presented. The solid solution stability is important in the MoS$_2$ structure because both CoS$_2$ and FeS$_2$ forms pyrite structures. This means that only concentrations below fifty percent are stable in the MMoS$_2$ structure before pyrite is formed. Therefore, as shown in Figure~\ref{fig:1T_turn_max} the results are depicted for only the stable compositions of MMoS$_2$ and not the pyrite phase.

First, it is known that the catalytic activity of the plane parallel to the lateral or horizontal axis(basal plane) of a pure 1T and 2H MoS$_{2}$ configurations mainly becomes apparent from the structure's affinity for binding hydrogen at surface S sites. Meaning that one would expect configurations comprised of more Mo atoms to have hydrogenation steps (associative: ${\Delta}$G5 to ${\Delta}$G6, ${\Delta}$G9 to ${\Delta}$G10, and dissociative: ${\Delta}$G15 to ${\Delta}$G16) which corresponds to the largest positive step that needs to be overcome from free energy formations of N$_{2}$ to N$_{2}$H (${\Delta}$G5 to ${\Delta}$G6), NH$_{2}$ to NH$_{3}$ (${\Delta}$G9 to ${\Delta}$G10), and N to NH (${\Delta}$G15 to ${\Delta}$G16). This can be seen in Figure\ref{fig:1T_step}, where more concentrations of Mo in the structure yields positive steps at the depicted formation energy regions (associative: ${\Delta}$G5 to ${\Delta}$G6, ${\Delta}$G9 to ${\Delta}$G10, and dissociative: ${\Delta}$G15 to ${\Delta}$G16). Also, one would expect structures to have hydrogen as a large positive step, due to internal bonding length of N-N dramatically increases as more hydrogen atoms are implemented in the species (Figure\ref{fig:ads}).

Second, nitrogen binds in a di-sigma bond (that is formed by overlapping between atomic orbitals) on the surface with metal atoms, and that Co and Fe atoms have a higher affinity to nitrogen than Mo atoms. Thus causing structures having more concentration of Fe and Co to have the positive determining step associated with braking N-N bonding (dissociative: ${\Delta}$G14 to ${\Delta}$G15 going from N$_{2}$ to N) because nitrogen binds very strongly to Fe and Co atoms compared to Mo. This can be seen when looking at lower concentrations of Mo atom in conjunction with a higher concentration of Fe and Co atoms in Figure\ref{fig:1T_step} of the dissociative mechanism going from N$_{2}$ to N (${\Delta}$G14 to ${\Delta}$G15).

Based on the two aforementioned ideas, one would expect a region where there is a shift between which positive step to be more dominant, either hydrogenation step or N-N bonding breaking step. In fact based on the dissociative mechanism illustrated by Figure~\ref{fig:1T_step}(A,C), one can see the largest step going from ${\Delta}$G14 to ${\Delta}$G15 (N$_{2}$ to N) to dramatically increase when more of Fe and or Co is implemented in the structure, because the strong di-sigma bonding with N-N and metal atoms. However, when the ratio of Mo to Fe/Co shifts from a structure containing more Mo atoms, the largest positive steps are affiliated with the hydrogenation as expressed by the associative mechanism of Figure~\ref{fig:1T_step}(B,D) going from ${\Delta}$G5 to ${\Delta}$G6 (N$_{2}$ to N$_{2}$H), ${\Delta}$G7 to ${\Delta}$G8 (N$_{2}$H$_{2}$ to N$_{2}$H$_{3}$), and ${\Delta}$G9 to ${\Delta}$G10 (NH$_{2}$ to NH$_{3}$) due to the first expressed idea of this section. To better illustrate this concept of shifting of positive steps and also to help determine the best structures for this mechanism, ternary plots of the maximum positive (Gibbs energy) step to overcome for a give reaction pathway is illustrated in Figure~\ref{fig:1T_turn_max}.

\subsection{Onset Maximum Positive Determining Step}
Figure\ref{fig:1T_turn_max}A and Figure\ref{fig:1T_turn_max}B illustrate the largest energy barrier that every composition for both morphologies must overcome. If this energy can be overcome by either an applied electric field in the case of electrocatalyst or a photogenerate potential in the case of photocatalyst, the reaction is free to evolve to completion. These values are in the absence of any kinetic considerations that may increase the required over-potential. The white area of the ternary plots corresponds to the formation of a pyrite structure. This is associated with a high concentration of Fe/Co to Mo atoms are incorporated in the structure. The points are associated with each DFT predictions. The contours are interpolation between the DFT predictions. It is reasoned that the colored region is a continuous design space and all compositions should be obtainable. The method of interpreting these plots is to recognize that Mo, Fe, and Co bound the space at the corners of the triangle. Along the edges of the triangle there is a binary composition of transition metals and within the triangle is a ternary composition. Also, it is important to note that these plots are only representing the largest positive step attained from Figure\ref{fig:1T_step} by taking the difference between each step and only recording maximum positive step. This step is noted as the largest energy barrier to overcome, therefore, the lowest value in Figure~\ref{fig:1T_turn_max} will be the best to implement because these structures will require the least amount of implemented potential. 
 
Based on findings mentioned in the previous section, one can visually confirm in Figure~\ref{fig:1T_turn_max} the shift in energy from a composition that has an affinity for hydrogenation to a composition with an affinity for N-N splitting. Where this shift is very clear for the dissociative mechanism of 1T and 2H structures, going from a pure Mo structure (bottom left corner) along the Fe/Co line. This region is saturated by higher energies toward the middle of the lower triangle for both the dissociative and associative mechanisms, but definitely more apparent in the dissociative mechanism. It is also noted that the highest energy for the associative 2H (Figure~\ref{fig:1T_turn_max}A) mechanism is attained for a pure Mo structure because the 2H structure has more Mo atoms and in turn, has higher affinity to H. The largest positive potential is associated with the hydrogenation, first point discussed in the previous section. This can be further convinced by the accept performance of 2H MoS$_2$ for HER. 

\begin{table}
\begin{center}
\begin{tabular}{ c |c |c }
\hline
 Pathway & 1T ($\eta$[eV]) & 2H ($\eta$[eV]) \\
 \hline
 Associative & Mo$_{0.75}$Co$_{0.25}$S$_{2}$ (1.23) & Mo$_{0.43}$Fe$_{0.57}$S$_{2}$ (1.14) \\
 Dissociative & Mo$_{0.75}$Fe$_{0.25}$S$_{2}$ (0.10) & Mo$_{0.43}$Fe$_{0.57}$S$_{2}$ (0.35) \\
 \hline
\end{tabular}
\end{center}
\caption{List of the best possible compositional points for each reaction pathway (associative and dissociative) and surface morphology (1T and 2H). These points correspond to the starred points in Figure~\ref{fig:1T_turn_max}. The associated over-potential determined is based on thermodynamic barrier energy and does not consider any kinetics. The best point overall is a Mo$_{0.75}$Fe$_{0.25}$S$_{2}$ 1T surface along the dissociative reaction pathway. }
\label{table:best_points}
\end{table}

Focusing on the minimum positive potential for the 1T and 2H associative (Figure~\ref{fig:1T_turn_max}(B,D)), where the associative mechanism tends to have more hydrogenation step to gradually weaken the N-N bonding till the very last steps. One would expect a structure having more Mo atoms and few Co/Fe atoms to be more dominant. This is seen for the 1T and 2H associative mechanisms, where the mechanism preferred Co atom due to the fact that Co is more electronegative than Fe. That is way higher concentration of Fe that is needed to satisfy the same energy levels as fewer Co implemented in the structure. Of course, this is not a unary or binary transition metal composition because the mechanisms still need to have hydrogenation and N-N bond breaking steps to fully form NH$_{2}$ from N$_{2}$. Therefore, structures have to have a balance of atomic components that both have an affinity for hydrogenation and N-N bond breaking. The structure that proves to have the best performance for the associative mechanism is a combination of 2Co and 6Mo atoms (Mo$_{0.75}$Co$_{0.25}$S$_{2}$) in the 8 metal positions for 1T structure and 3Mo and 4Fe atoms (Mo$_{0.43}$Fe$_{0.57}$S$_{2}$) in the 7 metal positions for the 2H structure that resulted in approximately 1.23 eV for the positive limiting step (over-potential ignoring kinetic considerations). The 1.23eV over-potential is within the band gap energy of MoS$_2$. Further investigation is required to determine the change in and gap energy for the compositional space.

Contrary to the associative mechanism, for the 1T and 2H dissociative (Figure~\ref{fig:1T_turn_max}) mechanisms, the N-N bonding is broken on the second reaction and the latter steps in the reaction are responsible for hydrogenations. Based on the claims for the associative mechanism, a structure with lower affinity to nitrogen will be associated with Fe-containing composition and reaction associated with a hydrogen containing species tend towards Mo incorporated compositions. This requires a balance structure also needs to effectively associate the hydrogenation that happens late in the reaction. The 1T dissociative mechanism preferred to have 6Mo and 2Fe atoms (Mo$_{0.75}$Fe$_{0.25}$S$_{2}$) in the 8 metal positions and the 2H preferred to have 4Fe and 3Mo atoms (Mo$_{0.43}$Fe$_{0.57}$S$_{2}$) in the 7 metal positions that resulted in approximately 1 eV for the positive determining step. It is also noted that these results should only be used as trends and not exact results because of approximation in the DFT method. The surface can be expected to be covered with hydrogen atom that forms a natural environment that can't be accounted for using the current approach. Also, the typical formation of hydrogen gas can end up being very fast unless the surface is covered with N adatoms rather than H adatoms, which contributes to possible discrepancies. 

\section{Conclusion}
In this study a DFT approach was implemented to explore the design space of MMoS$_2$ M=Mo,Fe,Co, which would otherwise be time and cost prohibitive from a pure experimental point of view. Moreover, the first principle approach allow additional physics to be understood that would otherwise be difficult to determine experimentally. The study limits the investigation to only the thermodynamic energy barriers for two reaction pathways in the absence of kinetic considerations. The study found that strong affinity of hydrogen species towards Mo and a similarly strong affinity of nitrogen towards Fe. The resulting structure was determined to be an alloy that combined both Mo and Fe. While both reactions paths could be optimized to obtain over-potentials below 2eV the most desired and the lowest of the pathways was the dissociative pathway. The optimal structure was a Mo$_{0.75}$Fe$_{0.25}$S$_{2}$ with a over-potential of approximately 0.1eV in the absence of kinetic consideration. An additional benefit of the 1T phase, which is associated with higher electonic mobility. Future studies will investigate the optics, electron mobility, and the intermediate transitions on this phase.

\footnotesize{
\bibliography{journal} 
\bibliographystyle{rsc} 
}


\end{document}